\documentclass[showpacs,twocolumn,preprintnumbers,showkeys,superscriptaddress,amsmath,amssymb,nofootinbib]{revtex4-2}

\usepackage{color}
\usepackage{amsmath}
\usepackage{amssymb}
\usepackage[utf8]{inputenc}
\usepackage{amsfonts}
\usepackage{bm}
\usepackage{bbold}
\usepackage{graphics}
\usepackage{graphicx}

\newcommand{\be}{\begin{equation}}
\newcommand{\ee}{\end{equation}}
\newcommand{\ba}{\begin{eqnarray}}
\newcommand{\ea}{\end{eqnarray}}

\def\ni{\noindent}

\begin{document}

%\linenumbers
%
%\begin{center}

\title{\Large A Dirac-material-inspired non-linear electrodynamic model}

\author{M. J. Neves} \email{mariojr@ufrrj.br}
%\affiliation{Department of Physics and Astronomy, University of Alabama, Tuscaloosa, Alabama 35487, USA}
\affiliation{Departamento de F\'{i}sica, Universidade Federal Rural do Rio de Janeiro, BR 465-07, 23890-971, Serop\'edica, RJ, Brazil}

\author{Patricio Gaete} \email{patricio.gaete@usm.cl}
\affiliation{Departamento de F\'{i}sica and Centro Cient\'{i}fico-Tecnol\'ogico de Valpara\'{i}so-CCTVal,
Universidad T\'{e}cnica Federico Santa Mar\'{i}a, Valpara\'{i}so, Chile}

\author{ L. P. R. Ospedal }  \email{ leoopr@cbpf.br}
\affiliation{Centro Brasileiro de Pesquisas F\'isicas, Rua Dr. Xavier Sigaud
150, Urca, Rio de Janeiro, Brasil, CEP 22290-180}

\author{J. A. Helay\"el-Neto}\email{helayel@cbpf.br}
\affiliation{Centro Brasileiro de Pesquisas F\'isicas, Rua Dr. Xavier Sigaud
150, Urca, Rio de Janeiro, Brasil, CEP 22290-180}

\date{\today}

\begin{abstract}
\ni
We propose and study the properties of a non-linear electrodynamics that emerges inspired on the physics of Dirac materials. This new electrodynamic model is an extension of the one-loop corrected non-linear effective Lagrangian computed in the work of ref. \cite{Keser}. In the particular regime of a strong magnetic and a weak electric field, it reduces to the photonic non-linear model worked out by the authors of ref. \cite{Keser}. We pursue our investigation of the proposed model by analyzing properties of the permittivity and permeability tensors, the energy- momentum tensor and wave propagation effects in presence of a uniform magnetic background. It is shown that the electrodynamics here presented exhibits the vacuum birefringence phenomenon. Subsequently, we calculate the lowest-order modifications to the interaction energy, considering still the presence of a uniform external magnetic field. Our analysis is carried out within the framework of the gauge-invariant but path-dependent variables formalism. The calculation reveals a screened Coulomb-like potential with an effective electric charge that runs with the external magnetic field but, as expected for Dirac-type materials, the screening disappears whenever the external magnetic field is switched off.
\end{abstract}

\pagestyle{myheadings}

\maketitle

\pagestyle{myheadings}
\markright{A Dirac-material-inspired non-linear electrodynamic model}

\section{Introduction}

As well-known, over the last decades, the area of quantum condensed matter physics and quantum field theory have become increasingly interconnected. This success is based on the fact that phenomena appearing in different systems are unified by common physical mechanisms \cite{Nagaosa,Tsvelik}. An illustrative example arises when one considers Weyl and Dirac semimetals, where it was shown in ref.  \cite{Keser} that quantum vacuum non-linear effects contribute to the experimentally observed high field magnetization. These effects have their origin in the well-known result due to Heisenberg and Euler \cite{Euler}.

In this context, it may be recalled that the subject of quantum vacuum  non-linearities
has been of great interest since the pioneering work by Euler and Heisenberg \cite{Euler}. In fact, in the case of non-linear electrodynamic models that appear as effective descriptions by summing up electron-positron quantum effects, the polarizable medium is the Quantum Electrodynamics (QED) vacuum, made up of virtual electrons and positrons. That is the case of the Euler-Heisenberg effective model. Incidentally, it is of interest to notice that Schwinger later confirmed this remarkable quantum prediction of light-by-light scattering from QED \cite{Schwinger}.

In this connection, we call attention to the fact that one of the striking physical effects of the Heisenberg and Euler result has been vacuum birefringence. In other words, the quantum vacuum when is stressed by external electromagnetic fields behaves as if it were a birefringent material medium, which has been emphasized from different perspectives \cite{Adler, Costantini, Ruffini, Dunne, Battesti, Sarazin}. However, this optical phenomenon has not yet been confirmed. Mention should be made, at this point, to recent results from ATLAS and CMS collaborations at the Large Hadron Collider (LHC), which have reported on the high energy gamma-gamma pair emission from virtual gamma-gamma scattering in ultraperipheral Pb-Pb collisions \cite{ATLAS, CMS,Schoeffel}. It is worthy to emphasize here that, in these results, there is no modification of the optical properties of the vacuum \cite{Robertson2}. We further mention that the coming of laser facilities has given rise to various proposals to probe quantum vacuum  non-linearities \cite{Battesti2, Ataman,Robertson2}.

Inspired by these observations, the purpose of this paper is to further elaborate on the physical content of the non-linear one-loop corrected effective electrodynamics discussed in ref. \cite{Keser}. Of special interest will be to consider the non-linear electrodynamics for Dirac materials in the familiar language of standard quantum field theory. Along these lines, we establish a connection between condensed matter and quantum field effective theories. Evidently, these connections are of interest in providing unifications among diverse models. We clarify that the model we present in this contribution is not derived as an effective physics stemming from Dirac materials. What we actually do is to write down a Lagrangian density that extends the effective physics worked out in ref. \cite{Keser} and recovers the latter in the special limit of a strong magnetic and weak electric field. Once the Dirac-material-inspired action is formulated, we obtain the field equations, conservation laws, as well as the properties of the medium due to non-linear contributions. Next, we study the propagation effects of this non-linear electrodynamics under an external and uniform magnetic field. We add the prescription of linearization expanding the Lagrangian density of the theory up to second order in the propagating fields. Thereby, we are able to extract the birefringence phenomenon in some Dirac materials. At last, to make contact with the discussion of the Coulomb screening that occurs in Dirac materials \cite{Wehling}, we calculate the interaction energy in Dirac materials under an uniform magnetic field via path-dependent variables formalism, and show that an effective charge comes out that depends on the external magnetic field.
The organization of the paper is as follows: In Section \ref{sec2}, we introduce a new non-linear electrodynamics for Dirac materials and analyze aspects such as equations of motion, the permittivity and permeability tensors and the corresponding energy-momentum tensor. In Section \ref{sec3}, we study wave propagation effects under a uniform magnetic background. In Section \ref{sec4}, we consider the birefringence phenomenon for this new electrodynamics. In Section \ref{sec5}, we compute the interaction energy for this electrodynamics under a uniform magnetic field. Our discussion is accomplished using the gauge-invariant but path-dependent variables formalism, which is an alternative to the usual Wilson loop approach. Finally, some concluding comments and considerations are cast in Section \ref{sec6}.
Throughout this work, we adopt natural units $\hbar=c=1$  with $4 \pi \epsilon_0 = 1$, and the Minkowski metric $\eta^{\mu\nu}=\mbox{diag}(+1,-1,-1,-1)$.
We use the conversion $1\,\mbox{m}=5 \times 10^{12} \, \mbox{MeV}^{-1}$. The electric and magnetic fields have squared-energy mass dimension, where the conversion of Volt/m and Tesla (T) to the natural system is given by $1 \, \mbox{Volt/m}=2.27 \times 10^{-24} \, \mbox{GeV}^2$ and $1 \, \mbox{T} =  6.8 \times 10^{-16} \, \mbox{GeV}^2$, respectively.

\section{A non-linear electrodynamics for Dirac materials}
\label{sec2}
As was shown in the work \cite{Keser}, the authors computed the non-linear
one-loop contribution for a Dirac material, given by the effective Lagrangian density:
\begin{eqnarray}\label{LDMapprox}
{\cal L}_{DM}\simeq
%\frac{\Delta}{24\pi^2\lambda_{D}^3}
\frac{\alpha_{D}}{24\pi^2} \left( \frac{\Delta^2}{ev_F} \right)^2
\left[ \, \frac{({\bf e}\cdot{\bf b})^{2}}{|{\bf b}|}+{\bf b}^2 \, \ln(|{\bf b}|) \, \right] \; ,
\end{eqnarray}
where $v_F$ denotes the Fermi velocity and $\Delta$ is the material gap with energy dimension.
This description holds in the regime of a strong magnetic field and a weak electric field. Furthermore, the dimensionless electromagnetic (EM) 
fields ${\bf e}$ and ${\bf b}$ depend on the fine structure constant $\alpha_{D}=e^2/v_F$ in the Dirac material,
\begin{eqnarray}\label{eDbD}
{\bf e}(\alpha_{D})=\frac{{\cal U} \, {\bf E}}{E_{\star}(\alpha_{D})}
\; , \;
{\bf b}(\alpha_{D})=\frac{{\cal U}^{-1} \, {\bf B}}{B_{\star}(\alpha_{D})} \; ,
\end{eqnarray}
with $E_{\star}(\alpha_{D})$ and $B_{\star}(\alpha_{D})$ being the critical Schwinger's electric and magnetic fields in the material, such that
\begin{eqnarray}\label{Estar}
E_{\star}^{2}(\alpha_{D}) = v_F^2 \,  B_{\star}^{2}(\alpha_{D})=\left( \frac{\Delta^2}{e v_F} \right)^2
%\simeq 4 \times 10^{12} \, \mbox{eV}/\mbox{m}^3
\; ,
\end{eqnarray}
and ${\cal U}$ is a symmetric matrix with $\det({\cal U})=1$. Thereby, the terms ${\cal U} \, {\bf E}$
and ${\cal U}^{-1} \, {\bf B}$ in (\ref{eDbD}) are linear combinations of ${\bf E}$ and ${\bf B}$, respectively.
%The physical parameters from (\ref{LDMapprox}) are resumed here as : $\Delta$ is the material gap with energy dimension and $\lambda_{D}=v/\Delta$
%is the Dirac wavelength in the material.
The Dirac magneton for the materials is $\mu_{D}=e\,v_F^2/(2\Delta)$, that replaces the Bohr magneton in the theory.
All these quantities defined previously depend on the nature of the material. Some particular cases are resumed
in the table \ref{table1}, where we have used the natural units (to make contact with the usually adopted units in condensed matter physics, we remind the reader that $1 \, \mbox{Volt/m}=2.27 \times 10^{-24} \, \mbox{GeV}^2$ and $1 \, \mbox{T} =  6.8 \times 10^{-16} \, \mbox{GeV}^2$ ). The Fermi velocity can be determined by the relation $v_{F}=\alpha/\alpha_{D}$ for each material.
%$\alpha_{D}=c\alpha/v\simeq 3$ is the fine structure constant in the material, in which $v=c/400=2 \times 10^{6}$ m/s is the Dirac velocity.
%

%
%\begin{table}
%
%\centering
%
%\begin{tabular}{|l|l|l|l|l|l|}
%\hline
%after \\: \hline or \cline{col1-col2} \cline{col3-col4} ...
%\quad & \quad $\Delta (\mbox{meV})$ \quad & \quad $\alpha_{D}/\alpha$ \quad & \quad $E_{\star}(\mbox{V/cm})$ \quad & \quad $B_{\star}(\mbox{mT})$ \quad \\
%\hline
%\hline
%\quad \mbox{QED} \quad & \quad $5\times10^{8}$ \quad & \quad $1$ \quad & \quad $1.3\times10^{16}$ \quad & \quad $4.4 \times 10^{12}$ \quad \\
%\hline
%\quad \mbox{PbSnTe} \quad & \quad \; $31.5$ \quad & \quad $580$ \quad & \quad $2.9\times 10^{4}$ \quad & \quad $5.6 \times 10^{3}$ \quad \\
%\hline
%\quad \mbox{BiSb} \quad & \quad \; $7.75$ \quad & \quad $188$ \quad & \quad $571$ \quad & \quad $36$ \quad \\
%\hline
%\quad \mbox{TaAs} \quad & \quad \quad $0$ \quad & \quad $357$ \quad & \quad $0$ \quad & \quad $0$ \quad  \\
%\hline
%\end{tabular}
%
%\caption{The comparison of the parameters for some materials and the QED. The fine structure constant in Dirac material $(\alpha_{D})$
%is defined in relation to QED $(\alpha)$, {\it i.e.}, $\alpha_{D}/\alpha=v^{-1}$.}\label{table1}
%
%\end{table}
%
%
\begin{table}
\centering
\begin{tabular}{|l|l|l|l|l|l|}
\hline
%after \\: \hline or \cline{col1-col2} \cline{col3-col4} ...
& $\Delta (\mbox{meV})$ & $v_{F}=\alpha/\alpha_{D}$ & $E_{\star}(\mbox{eV}^2)$ & $B_{\star}(\mbox{eV}^2)$ \\
\hline
\hline
\mbox{QED} & $5\times10^{8}$ & $1$ & $2.95 \times 10^{12}$ & $3.0 \times 10^{12}$ \\
\hline
\mbox{PbSnTe} & $31.5$ & $0.0017$ & $6.58$ & $3.8\times 10^{-3}$ \\
\hline
\mbox{BiSb} & $7.75$ & $0.0053$ & $0.13$ & $24.4$ \\
\hline
\mbox{TaAs} & $0$ & $0.0028$ & $0$ & $0$  \\
\hline
\end{tabular}
\caption{The comparison of the parameters for some materials and the QED. The fine structure constant in Dirac material $(\alpha_{D})$
is defined in relation to QED $(\alpha)$, {\it i.e.}, $\alpha_{D}/\alpha=v_{F}^{-1}$.}\label{table1}
\end{table}
%

%
%$\Delta=50$ meV is the material gap with energy dimension, $\lambda_{D}=\hbar v/\Delta=10$ nm is the Dirac wavelength in the material, $\alpha_{D}=c\alpha/v\simeq 3$ is the fine structure constant in the material, in which $v=c/400=2 \times 10^{6}$ m/s is the Dirac velocity.
%

%
%{\color{red} The Lagrangian density (\ref{LDMapprox}) can be rewritten in terms of the fields ${\bf E}$ and ${\bf B}$.
%Using the relations (\ref{Estar}), we obtain
%
%\begin{eqnarray}\label{LDMapproxEB}
%{\cal L}_{DM}\simeq
%%\frac{\Delta}{24\pi^2\lambda_{D}^3}
%\frac{\alpha_{D}}{24\pi^2}
%\left[ \, \frac{v}{\beta} \, \frac{({\bf E}\cdot{\bf B})^{2}}{|{\bf B}|}
%+ v^2 \, {\bf B}^2 \, \ln\left(v \, \frac{|{\bf B}|}{\beta}\right) \, \right] \; ,%
%\end{eqnarray}
%
%where the $\beta$-parameter is identified as the Schwinger's electric field in
%the material $\beta := E_{\star}(\alpha_{D})$ defined in (\ref{Estar}), and the magnitude
%of the fields are such that $|{\bf B}| \approx \beta$, with $|{\bf B}| \gg |{\bf E}|$.
%
%\begin{eqnarray}\label{beta}
%\beta=\sqrt{ \frac{\Delta}{\alpha_{D}\,\lambda_{D}^3} }=\frac{\Delta^2}{e v} \; .
%\end{eqnarray}
%
%}
%

\begin{figure}[th]
%\vspace{-5pt}
\centering
\includegraphics[width=0.48\textwidth]{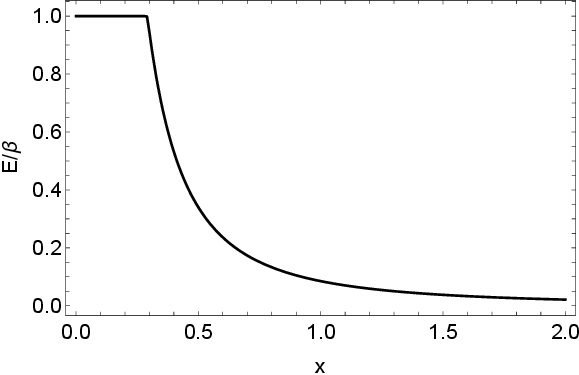}
%\quad\quad
%\includegraphics[width=0.45\textwidth]{n2x.eps}
\caption{The dimensionless variable $E/\beta$ versus $x=r\sqrt{\beta}$ for the QED.}
\label{Evsx}
\end{figure}

%
We propose a general non-linear electrodynamics (ED) governed by the Lagrangian density
\begin{eqnarray}\label{LDM}
{\cal L}_{DM}=- \frac{\alpha_{D}}{24\pi^2} \, {\cal F} \, \ln\left[\,1-\frac{2\,{\cal F}}{\beta^2}+\frac{2 \, {\cal G}^2}{\beta^3\sqrt{-2\,{\cal F}}}\, \right] \; ,
\end{eqnarray}
where the $\beta$-parameter is identified as the Schwinger's electric field in
the material, namely, $\beta := E_{\star}(\alpha_{D})$. Moreover, ${\cal F}$ and ${\cal G}$ are the Lorentz and gauge-invariants
\begin{subequations}
\begin{eqnarray}
{\cal F}\!&=&\!-\frac{1}{4} \, F_{\mu\nu}^{2}=\frac{1}{2} \, \left( \, {\bf E}^2-{\bf B}^2 \, \right) \; ,
\label{invF}
\\
%\hspace{0.2cm} \mbox{and} \hspace{0.2cm}
{\cal G}\!&=&\!-\frac{1}{4} \, F_{\mu\nu}\widetilde{F}^{\mu\nu}={\bf E}\cdot{\bf B} \; ,
\label{invG}
\end{eqnarray}
\end{subequations}
in which $F_{\mu\nu}$ denotes the field strength tensor and
$\widetilde{F}^{\mu\nu}=\epsilon^{\mu\nu\alpha\beta}F_{\alpha\beta}/2$ corresponds to the dual tensor,
while ${\bf E}$ and ${\bf B}$ are the electric and magnetic fields, respectively.
Using the approximation $|{\bf B}| \gg |{\bf E}|$ and the relations (\ref{Estar}), one can show that 
the Lagrangian density (\ref{LDM}) reduces to the expression (\ref{LDMapprox}).
If the magnitude of the magnetic field is $|{\bf B}|>|{\bf E}|$, the non-linear Lagrangian density
(\ref{LDM}) is real. Otherwise, if $|{\bf E}| > |{\bf B}|$, this theory can present complex terms
in the logarithmic argument. Notice also that eq. (\ref{LDM}) goes to zero in the limit $|{\bf E}| \rightarrow |{\bf B}|$.
It should be mentioned that the effective Lagrangian (\ref{LDMapprox})
is valid for $\Delta \neq 0$, which implies into $\beta \neq 0$ by the relation (\ref{Estar}). Thus,
the Lagrangian density (\ref{LDM}) is valid for $\beta \neq 0$ with any magnitude of the electromagnetic field,
and the non-linear effects disappear when $\beta \rightarrow \infty$. 
We add the usual Maxwell Lagrangian to the non-linear
electrodynamics in eq. (\ref{LDM}) :
\begin{eqnarray}\label{Lmodel}
{\cal L}={\cal F}-\frac{\alpha_{D}}{24\pi^2} \, {\cal F} \, \ln\left[\,1-\frac{2\,{\cal F}}{\beta^2}+\frac{2 \, {\cal G}^2}{\beta^3\sqrt{-2\,{\cal F}}}\, \right] \; ,
\end{eqnarray}
in which the usual Maxwell theory is recovered in the limit $\beta \rightarrow \infty$. Therefore, we have a non-linear ED defined in the Minkowski space-time, whose the Dirac material is spatial $3$D. The action principle applied to eq. (\ref{Lmodel}) yields the field equation
\begin{equation}\label{eq}
\partial_{\mu}\left[ \, \xi({\cal F},{\cal G}) \, F^{\mu\nu}+\chi({\cal F},{\cal G}) \, \widetilde{F}^{\mu\nu} \, \right]=0 \; ,
\end{equation}
where the coefficients $\xi({\cal F},{\cal G})$ and $\chi({\cal F},{\cal G})$ are functions of ${\bf E}$ and ${\bf B}$,
according to
\begin{subequations}
\begin{eqnarray}
\xi({\cal F},{\cal G}) \!&=&\! 1+\frac{\alpha_{D}}{48\pi^2}\frac{ \sqrt{2} \, {\cal G}^2+4\beta \, \sqrt{-{\cal F}} \, {\cal F} }{ \sqrt{2} \, {\cal G}^2+\sqrt{-{\cal F}} \beta(-2{\cal F}+\beta^2) }
\nonumber \\
&&
\hspace{-0.5cm}
-\frac{\alpha_{D}}{24\pi^2} \, \ln\left[\,1-\frac{2\,{\cal F}}{\beta^2}+\frac{2 \, {\cal G}^2}{\beta^3\sqrt{-2\,{\cal F}}}\, \right]
 \; , \; \; \;
\\
\chi({\cal F},{\cal G}) \!&=&\! -\frac{\alpha_{D}}{6\pi^2}\frac{{\cal F} \, {\cal G}}{2\,{\cal G}^2+\sqrt{-2{\cal F}} \beta(-{\cal F}+\beta^2) }  \; .
\end{eqnarray}
\end{subequations}
As in the usual case, the dual tensor $\widetilde{F}^{\mu\nu}=\epsilon^{\mu\nu\alpha\beta}F_{\alpha\beta}/2$ satisfies the
Bianchi identity $\partial_{\mu}\widetilde{F}^{\mu\nu}=0$.
The modified Maxwell equations can be written in terms of the constitutive vectors,
\begin{subequations}
\begin{eqnarray}
\nabla\cdot{\bf D} \!&=&\! 0
\hspace{0.3cm} , \hspace{0.3cm}
\nabla\times{\bf E}+\frac{\partial{\bf B}}{\partial t} = {\bf 0} \; ,
\label{eqdivD}
\\
\nabla\cdot{\bf B} \!&=&\!0
\hspace{0.3cm} , \hspace{0.3cm}
\nabla\times{\bf H}-\frac{\partial{\bf D}}{\partial t} = {\bf 0} \; ,
\label{eqdivB}
\end{eqnarray}
\end{subequations}
where the components of ${\bf D}$ and ${\bf H}$ vectors are defined by $D_{i}=\epsilon_{ij}({\bf E},{\bf B})\,E_{j}$
and $H_{i}=(\mu_{ij})^{-1}({\bf E},{\bf B})\,B_{j}$, respectively. The permittivity symmetric tensor $\epsilon_{ij}$
and the inverse of the permeability tensor $(\mu_{ij})^{-1}$ are read below :
\begin{subequations}
\begin{eqnarray}
\epsilon_{ij}({\bf E},{\bf B}) \!&=&\! \left\{ 1 + \frac{\alpha_{D}}{48\pi^2}\frac{ (\sqrt{2} \, {\cal G}^2+4\beta \, \sqrt{-{\cal F}} \, {\cal F}) }{ \sqrt{2} \, {\cal G}^2+\sqrt{-{\cal F}} \beta(-2{\cal F}+\beta^2) }
\right.
\nonumber \\
&&
\hspace{-0.5cm}
\left.
-\frac{\alpha_{D}}{24\pi^2} \, \ln\left[\,1-\frac{2\,{\cal F}}{\beta^2}+\frac{2 \, {\cal G}^2}{\beta^3\sqrt{-2\,{\cal F}}}\, \right] \right\} \, \delta_{ij}
\nonumber \\
&&
\hspace{-0.5cm}
-\frac{\alpha_{D}}{6\pi^2}\frac{{\cal F} \, B_{i} \, B_{j} }{2\,{\cal G}^2+\sqrt{-2{\cal F}} \beta(-{\cal F}+\beta^2) } \; ,
\label{epsilonij}
\\
(\mu_{ij})^{-1}({\bf E},{\bf B}) \!&=&\! \left\{ 1 + \frac{\alpha_{D}}{48\pi^2}\frac{ (\sqrt{2} \, {\cal G}^2+4\beta \, \sqrt{-{\cal F}} \, {\cal F}) }{ \sqrt{2} \, {\cal G}^2+\sqrt{-{\cal F}} \beta(-2{\cal F}+\beta^2) }
\right.
\nonumber \\
&&
\hspace{-0.5cm}
\left.
-\frac{\alpha_{D}}{24\pi^2} \, \ln\left[\,1-\frac{2\,{\cal F}}{\beta^2}+\frac{2 \, {\cal G}^2}{\beta^3\sqrt{-2\,{\cal F}}}\, \right] \right\} \, \delta_{ij}
\nonumber \\
&&
\hspace{-0.5cm}
+ \frac{\alpha_{D}}{6\pi^2}\frac{{\cal F} \, E_{i} \, E_{j} }{2\,{\cal G}^2+\sqrt{-2{\cal F}} \beta(-{\cal F}+\beta^2) } \; .
\label{muinv}
\end{eqnarray}
\end{subequations}
The inverse of the matrix (\ref{muinv}) yields the permeability tensor
\begin{eqnarray}
\mu_{ij}=\frac{1}{1+a}\left[ \, \delta_{ij} - \frac{b \, E_{i} \, E_{j} }{1+a+b \, {\bf E}^2} \, \right] \; ,
\end{eqnarray}
in which, for convenience, we define
\begin{subequations}
\begin{eqnarray}
a \!&=&\! \frac{\alpha_{D}}{48\pi^2} \frac{ \sqrt{2} \, {\cal G}^2+4\beta \, \sqrt{-{\cal F}} \, {\cal F} }{ \sqrt{2} \, {\cal G}^2+\beta\sqrt{-{\cal F}} (-2{\cal F}+\beta^2) }
\nonumber \\
&&
\hspace{-0.5cm}
-\frac{\alpha_{D}}{24\pi^2} \, \ln\left[\,1-\frac{2\,{\cal F}}{\beta^2}+\frac{2 \, {\cal G}^2}{\beta^3\sqrt{-2\,{\cal F}}}\, \right] \; ,
\\
b \!&=&\! \frac{\alpha_{D}}{6\pi^2} \frac{ {\cal F} }{2\,{\cal G}^2+\beta\sqrt{-2{\cal F}} (-{\cal F}+\beta^2) } \; .
\end{eqnarray}
\end{subequations}
The Lagrangian density of eq. (\ref{LDM}) yields the electric permittivity tensor presented in eqs. (\ref{epsilonij}) and (\ref{muinv}). It has two distinct eigenvalues, $\lambda_1$  and  $\lambda_2$ , which read:  $\lambda_1 = 1+a$ (with degree of degeneracy 2), whereas $\lambda_2 = 1+a-b\,{\bf B}^2$. By analyzing the positivity of the permittivity eigenvalues, and considering the situation such that the Dirac material shares an interface with some other material, dielectric or metal, we can pursue an investigation as to whether surface plasmons may show up in the interface. This inspection may be relevant if one has in mind to study whether polaritons may arise from the interaction between the (surface) plasmons in the interface and light. This is a potential application of the Lagrangian density (\ref{LDM}).
%
%In the optical basis of the EM field, the permittivity and permeability tensors can be diagonalized by a $SO(3)$
%transformation in which the diagonal elements are given by the corresponding eigenvalues :
%
%\begin{subequations}
%\begin{eqnarray}
%\lambda_{1\epsilon} &=& \lambda_{2\epsilon} = 1+a
%\; , \;
%\lambda_{3\epsilon} = 1+a-b\,{\bf B}^2 \; ,
%\label{eigenvaluesepsilon}
%\\
%\lambda_{1\mu} &=& \lambda_{2\mu}=\frac{1}{1+a}
%\; , \;
%\lambda_{3\mu} = \frac{1}{1+a+b\,{\bf E}^2} \; .
%\label{eigenvaluesmu}
%\end{eqnarray}
%\end{subequations}
%
Thereby, the positive conditions for the eigenvalues $1+a>0$, $1+a-b\,{\bf B}^2>0$
and $1+a+b\,{\bf E}^2>0$ insure that the previous tensors are positive.
The energy-momentum tensor associated with the equations (\ref{eqdivD}) and (\ref{eqdivB}) reads as below:
\begin{eqnarray}\label{tensorEnMomentum}
\Theta^{\mu\rho}=
\xi({\cal F},{\cal G}) \, F^{\mu\nu} F_{\nu}^{\;\;\rho}
\,+\,\eta^{\mu\rho} \left[ \, \chi({\cal F},{\cal G}) \, {\cal G}-{\cal L} \, \right] \; .
\end{eqnarray}
that satisfies the conservation law $\partial_{\mu}\Theta^{\mu\rho}=0$. The components of the
energy-momentum tensor are read
\begin{subequations}
\begin{eqnarray}
\Theta^{00}\!&=&\! \frac{1}{2} \left( {\bf E}^{\,2} + {\bf B}^2 \right)
+ \frac{\alpha_{D}}{48\pi^2}\frac{ \left( \sqrt{2} \, {\cal G}^2+4\beta \, \sqrt{-{\cal F}} \, {\cal F} \right) {\bf E}^{\,2}
}{ \sqrt{2} \, {\cal G}^2+\sqrt{-{\cal F}} \beta(-2{\cal F}+\beta^2) }
\nonumber \\
&&
\hspace{-0.5cm}
-\frac{\alpha_{D}}{24\pi^2} \, \ln\left[\,1-\frac{2\,{\cal F}}{\beta^2}+\frac{2 \, {\cal G}^2}{\beta^3\sqrt{-2\,{\cal F}}}\, \right] {\bf E}^2
\nonumber \\
&&
\hspace{-0.5cm}
-\frac{\alpha_{D}}{6\pi^2}\frac{{\cal F} \, ({\bf E}\cdot{\bf B})^2  }{2\,{\cal G}^2+\sqrt{-2{\cal F}} \beta(-{\cal F}+\beta^2) } \; ,
\label{Thetacomp00}
\hspace{-0.5cm}
\\
\Theta^{0i}\!\!&=&\!\! \left\{ 1+\frac{\alpha_{D}}{48\pi^2}\frac{ \sqrt{2} \, {\cal G}^2+4\beta \, \sqrt{-{\cal F}} \, {\cal F} }{ \sqrt{2} \, {\cal G}^2+\sqrt{-{\cal F}} \beta(-2{\cal F}+\beta^2) }
\right.
\nonumber \\
&&
\hspace{-0.5cm}
\left.
-\frac{\alpha_{D}}{24\pi^2} \, \ln\left[\,1-\frac{2\,{\cal F}}{\beta^2}+\frac{2 \, {\cal G}^2}{\beta^3\sqrt{-2\,{\cal F}}}\, \right]
\right\} \left( {\bf E} \times {\bf B} \right)^{i} . \;\;\;
\label{Thetacomp0i}
%\hspace{-0.5cm}
%\nonumber \\
\end{eqnarray}
\end{subequations}
The result (\ref{Thetacomp00}) requires that $|{\bf B}|>|{\bf E}|$, thus the energy density and the
Poynting vector are real. The same condition must be imposed for the eigenvalues $\lambda_1 = 1 + a$
and $\lambda_2 = 1 + a - b\,{\bf B}^2$ to be real. All the results of the Maxwell ED are recovered in the limit
$\beta \rightarrow \infty$.
In the electrostatic case, we consider ${\bf B}={\bf 0}$ in the previous equations.
For a point-like charge, $\rho({\bf r})=Q \, \delta({\bf r})$, and the Gauss law yields
${\bf D}=Q \, \hat{{\bf r}}/r^2$. Using the permittivity tensor for ${\bf B}={\bf 0}$, the
electrostatic field $(E)$ satisfies the following equation :
\begin{eqnarray}\label{Er}
E\left[ \, 1+\frac{\alpha_{D}}{24\pi^2}\frac{E^2}{\beta^2-E^2}-\frac{\alpha_{D}}{24\pi^2} \, \ln\left(1-\frac{E^2}{\beta^2}\right)  \, \right]=\frac{Q}{r^2} \; . \hspace{0.4cm}
\end{eqnarray}
Notice that the limit $\beta \rightarrow \infty$ reduces (\ref{Er}) to the usual electrostatic field of a point-like charge. We solve numerically the equation (\ref{Er}) to obtain the plot (\ref{Evsx}). The figure shows the
dimensionless variable $E/\beta$ (electric field for point-like charge over the $\beta$-parameter)
versus $x=r\,\sqrt{\beta}$ (radial distance times the squared root of $\beta$-parameter)
for the QED. In this plot, we use the fundamental charge in natural units
$Q=e=0.085$, and $\alpha_{D}=\alpha=1/137=0.0072$.
\noindent

The electric field is finite at the origin $(r \rightarrow 0)$,
and it is given by the critical electric field of the material, {\it i. e.},
$E(r=0)=\beta=2.95 \times 10^{12} \, \mbox{eV}^2=2.95\,\mbox{MeV}^2$ (in the case of QED).
%This behavior is similar for the BiSb.
The energy density in the electrostatic case is
\begin{equation}
u_{E}=\frac{1}{2} \, {\bf E}^2\left[1+\frac{\alpha_{D}}{12\pi^2}\frac{{\bf E}^2}{ \beta^2-{\bf E}^2}-\frac{\alpha_{D}}{12\pi^2}\,\ln\left(1-\frac{{\bf E}^2}{\beta^2} \right) \right] \, , \;\;\;
\end{equation}
that is positive if $\beta > |{\bf E}|$, and the electrostatic energy stored in the electron
is
\begin{eqnarray}\label{Uint}
U=2\pi \int_{\lambda_c}^{\infty} \! r^2 \, dr \, \times
\nonumber \\
{\bf E}^2 \left[1+\frac{\alpha_{D}}{12\pi^2}\frac{{\bf E}^2}{ \beta^2-{\bf E}^2}-\frac{\alpha_{D}}{12\pi^2}\,\ln\left(1-\frac{{\bf E}^2}{\beta^2} \right) \right] \; ,
\end{eqnarray}
where $\lambda_c=3.86 \times 10^{-13} \, \mbox{m}=1.93 \, \mbox{MeV}^{-1}$ is the electron's Compton wavelength.
%the distance correspondent to the values of
%$E$ near from the critical field $\beta=2.95 \, \mbox{MeV}^2$
Using the expression of the electric field for a like-point charge in (\ref{Er}), and that $\alpha_{D}=\alpha=1/137$ for the QED,
we solve numerically the integral (\ref{Uint}) to obtain the finite electron's self-energy
%
%\begin{subequations}
\begin{eqnarray}
U = 0.023 \, \mbox{MeV}
%\;\; \mbox{for} \;\; \mbox{QED}
\; .
%\\
%U \!\!&=&\!\! 1.065 \, \mbox{eV} \;\; \mbox{for} \;\; \mbox{PbSnTe} \; ,
%\\
%U \!\!&=&\!\! 0.15 \, \mbox{eV} \;\; \mbox{for} \;\; \mbox{BiSb} \; .
\end{eqnarray}
%\end{subequations}
%

%
\section{Wave propagation properties in a uniform magnetic background}
\label{sec3}
We now examine wave propagation effects under a uniform magnetic field for the
non-linear ED (\ref{Lmodel}). Follo\-wing a similar procedure of ref. \cite{MJNevesEDN},
we start off our analysis by introducing a prescription for the external and uniform magnetic field through the gauge 4-potential $A_{\mu}=a_{\mu}+A_{B\mu}$, where $a_{\mu}$ is interpreted as the propagation $4$-potential,
and $A_{B\mu}$ is the potential associated with the external and uniform magnetic field ${\bf B}_{0}$.
Consequently, the EM field-strength tensor is also decomposed as $F_{\mu\nu}=f_{\mu\nu}+F_{B\mu\nu}$, in which $f^{\mu\nu}=\partial^{\mu}a^{\nu}-\partial^{\nu}a^{\mu}=\left( \, -e^{i} \, , \, -\epsilon^{ijk} \, b^{k} \, \right)$ denotes the EM field strength tensor of the propagating fields, whereas $F_{B}^{\;\,\mu\nu}=\partial^{\mu}A_{B}^{\;\,\nu}-\partial^{\nu}A_{B}^{\;\,\mu} =\left( \, 0 \, , \, -\epsilon^{ijk} \, B_{0}^{\,\,k} \, \right)$ corresponds to the field strength of the magnetic background. Using this prescription, the Lagrangian density (\ref{Lmodel}) up to second order in the propagation gauge field leads to
\begin{equation}\label{Lmodel4}
{\cal L}_{model}^{(2)} = -\frac{1}{4} \, c_{1} \, f_{\mu\nu}^{\, 2}
-\frac{1}{4} \, c_{2} \, f_{\mu\nu} \, \widetilde{f}^{\mu\nu}
%-\frac{1}{2} \, G_{0\mu\nu} \, f^{\mu\nu}
%\nonumber \\
%&&
%\hspace{-0.6cm}
+\,\frac{1}{8} \, Q_{B\mu\nu\kappa\lambda} \, f^{\mu\nu} \, f^{\kappa\lambda}
%+\frac{1}{8} \, R_{0\mu\nu\kappa\lambda\rho\sigma} \, f^{\mu\nu} \, f^{\kappa\lambda} \, f^{\rho\sigma}
%\nonumber \\
%&&
%\hspace{-0.6cm}
%+\frac{1}{16} \, S_{0\mu\nu\kappa\lambda\rho\sigma\omega\tau} \, f^{\mu\nu} \, f^{\kappa\lambda} \, f^{\rho\sigma} \, f^{\omega\tau}
\; ,
\end{equation}
where $\widetilde{f}_{\mu\nu}=\varepsilon_{\mu\nu\alpha\beta}f^{\alpha\beta}/2$  is the propagation field strength dual tensor. 
In addition, we have defined the tensor $Q_{B\mu\nu\kappa\lambda} $ at the background fields $(B)$ as follows
%
%\begin{widetext}
\begin{eqnarray}
%G_{0\mu\nu} \!&=&\! c_{1} \, F_{0\mu\nu}+ c_{2} \, \widetilde{F}_{0\mu\nu} \; ,
%\nonumber \\
Q_{B\mu\nu\kappa\lambda} \!&=&\! d_{1} \, F_{B\mu\nu}F_{B\kappa\lambda}
+d_{2} \, \widetilde{F}_{B\mu\nu}\widetilde{F}_{B\kappa\lambda}
\nonumber \\
&&
\hspace{-0.3cm}
+\,d_{3} \, F_{B\mu\nu}\widetilde{F}_{B\kappa\lambda}
+ d_{3} \, \widetilde{F}_{B\mu\nu} F_{B\kappa\lambda} \; ,
\end{eqnarray}
%\end{widetext}
%/
with $\widetilde{F}_{B\mu\nu}=\varepsilon_{\mu\nu\alpha\beta}\,F_{B}^{\;\,\,\alpha\beta}/2$ being the dual tensor of the background field.
The coefficients of the expansion $c_{i} \, (i=1,2)$ and $d_{i} \, (i=1,2,3)$
%$M_{i} \, (i=1,2,3,4)$ and $N_{i} \, (i=1,2,3,4,5)$
are given by
\begin{eqnarray}\label{coefficientsMN}
c_{1}=\left.\frac{\partial{\cal L}}{\partial{\cal F}}\right|_{{\bf B}_{0}} ,
\left. c_{2}=\frac{\partial{\cal L}}{\partial{\cal G}}\right|_{{\bf B}_{0}} ,
%\nonumber \\
\left. d_{1}=\frac{\partial^2{\cal L}}{\partial{\cal F}^2}\right|_{{\bf B}_{0}} ,
\nonumber \\
\left. d_{2}=\frac{\partial^2{\cal L}}{\partial{\cal G}^2}\right|_{{\bf B}_{0}} ,
%\nonumber \\
\left. d_{3}=\frac{\partial^2{\cal L}}{\partial{\cal F}\partial{\cal G}}\right|_{{\bf B}_{0}} ,
%\nonumber \\
%M_{1}=\left.\frac{\partial^3{\cal L}}{\partial{\cal F}^3}\right|_{{\bf B}_{0}}
%\, , \,
%\nonumber \\
%\left. M_{2}=\frac{\partial^3{\cal L}}{\partial{\cal G}^3}\right|_{{\bf B}_{0}} ,
%\nonumber \\
%\left. M_{3}=\frac{\partial^3{\cal L}}{\partial{\cal F}^2\partial{\cal G}}\right|_{{\bf B}_{0}} ,
%
%\left. M_{4}=\frac{\partial^3{\cal L}}{\partial{\cal F}\partial{\cal G}^2}\right|_{{\bf B}_{0}} ,
%
%\nonumber \\
%\left. N_{1}=\frac{\partial^4{\cal L}}{\partial{\cal F}^4}\right|_{{\bf B}_{0}} ,
%
%\left. N_{2}=\frac{\partial^4{\cal L}}{\partial{\cal G}^4}\right|_{{\bf B}_{0}} ,
%\nonumber \\
%\left. N_{3}=\frac{\partial^4{\cal L}}{\partial{\cal F}^3\partial{\cal G}}\right|_{{\bf B}_{0}} ,
%\nonumber \\
%\left. N_{4}=\frac{\partial^4{\cal L}}{\partial{\cal F}^2\partial{\cal G}^2}\right|_{{\bf B}_{0}} ,
%\nonumber \\
%\left. N_{5}=\frac{\partial^4{\cal L}}{\partial{\cal F}\partial{\cal G}^3}\right|_{{\bf B}_{0}} \, .
%\hspace{0.4cm}
\end{eqnarray}
For the particular model in eq. (\ref{Lmodel}), we obtain the coefficients in a magnetic background :
\begin{eqnarray}\label{coeficientsresult}
c_{1} &=& 1-\frac{\alpha_{D}}{24\pi^2}\frac{ {\bf B}_{0}^2}{\beta^2+{\bf B}_{0}^2}-\frac{\alpha_{D}}{24\pi^2} \ln \left(1+\frac{{\bf B}_{0}^2}{\beta^2}\right) \; ,
\nonumber \\
c_{2} &=& 0
\; , \;
d_{1}=\frac{\alpha_{D}}{12\pi^2} \frac{ {\bf B}_{0}^2+2\beta^2}{({\bf B}_{0}^2+\beta^2)^2}
\; ,
\nonumber \\
d_{2} &=& \frac{\alpha_{D}}{24\pi^2}\frac{|{\bf B}_{0}|/\beta}{ {\bf B}_{0}^2+\beta^2 }
\; , \;
d_{3}=0 \; .
%\nonumber \\
%M_{1}\!\!&=&\!\!\frac{\alpha_{D}}{6\pi^2} \frac{{\bf B}_{0}^2+3\beta^2}{({\bf B}_{0}^2+\beta^2)^3}
%\; , \;
%M_{2}=0 \; ,
%\nonumber \\
%M_{3}\!\!&=&\!\! 0
%\; , \;
%M_{4}=\frac{\alpha_{D}}{12\pi^2} \frac{ {\bf B}_{0}^2-\beta^2 }{\beta|{\bf B}_0|({\bf B}_{0}^2+\beta^2)^2} \; ,
%\nonumber \\
%N_{1} \!\!&=&\!\!\frac{2\alpha_{D}}{3\pi^2} \frac{ {\bf B}_{0}^2+4\beta^2 }{({\bf B}_{0}^2+\beta^2)^4}
%\; , \;
%N_{2}=-\frac{\alpha_{D}}{\pi^2}\frac{1}{\beta^2({\bf B}_0^2+\beta^2)^2} \; ,
%\nonumber \\
%N_{3} \!\!&=&\!\! 0 \; , \;
%N_{4} =\frac{\alpha_{D}}{12\pi^2}\frac{3{\bf B}_{0}^4-6{\bf B}_{0}^2 \, \beta^2-\beta^4}{{\bf B}_{0}^3 \, \beta \left({\bf B}_{0}^2+\beta^2\right)^3}
%\; , \;
%N_{5}=0 \; .
\hspace{0.7cm}
\end{eqnarray}
Taking into account the non-null coefficients, the Lagrangian density (\ref{Lmodel4}) can be recast as
\begin{eqnarray}\label{L2}
{\cal L}^{(2)}=-\frac{1}{4} \, c_{1} \, f_{\mu\nu}^{\, 2}
+\frac{d_1}{8} \, (F_{B\mu\nu}f^{\mu\nu})^{2}
+\frac{d_2}{8} \, (\widetilde{F}_{B\mu\nu}f^{\mu\nu})^{2} .
\;\;\;
\nonumber \\
\end{eqnarray}
%
%\begin{eqnarray}\label{L3}
%{\cal L}^{(3)}=\frac{d_1}{16} \, (F_{0\mu\nu}f^{\mu\nu}) \, f_{\rho\sigma}^2
%+\frac{d_2}{8} (\widetilde{F}_{0\mu\nu}f^{\mu\nu}) (\widetilde{f}_{\rho\sigma}f^{\rho\sigma})
%\nonumber \\
%-\frac{M_{1}}{16} \, (F_{0\mu\nu}f^{\mu\nu})^{3}
%-\frac{M_{4}}{16} \, (F_{0\mu\nu}f^{\mu\nu})(\widetilde{F}_{0\rho\sigma}f^{\rho\sigma})^{2}, \nonumber\\
%\end{eqnarray}
%
%\begin{eqnarray}\label{L4}
%{\cal L}^{(4)} \!&=&\! \frac{d_1}{32} \, (f_{\mu\nu}^{2})^{2}
%+\frac{d_2}{32} \, (f_{\mu\nu}\widetilde{f}^{\mu\nu})^2
%-\frac{M_{1}}{32} \, (F_{0\mu\nu}f^{\mu\nu})^{2} f_{\rho\sigma}^2
%\nonumber \\
%&&
%\hspace{-0.5cm}
%-\frac{M_{4}}{16} (F_{0\mu\nu}f^{\mu\nu})(\widetilde{F}_{0\kappa\lambda}f^{\kappa\lambda})(f_{\rho\sigma}\widetilde{f}^{\rho\sigma})
%\nonumber \\
%&&
%\hspace{-0.5cm}
%-\frac{M_{4}}{32} \, f_{\mu\nu}^2(\widetilde{F}_{0\kappa\lambda}f^{\kappa\lambda})^2
%+\frac{N_{1}}{384} \, (F_{0\mu\nu}f^{\mu\nu})^{4}
%\nonumber \\
%&&
%\hspace{-0.5cm}
%+\frac{N_2}{384} \, (\widetilde{F}_{0\mu\nu}f^{\mu\nu})^4
%+\frac{N_4}{64} \, (F_{0\mu\nu}f^{\mu\nu})^2 (\widetilde{F}_{0\rho\sigma}f^{\rho\sigma})^{2} .
%\;\;\;\hspace{0.5cm}
%\end{eqnarray}
%

%
The action principle applied to this Lagrangian density yields the following field equation
\begin{equation}\label{eqf}
\partial^{\mu}\left[ c_{1}\,f_{\mu\nu}
%+c_{2}\,\widetilde{f}_{\mu\nu}
-\frac{d_1}{2} \, F_{B\mu\nu} (F_{B\kappa\lambda}f^{\kappa\lambda})
-\frac{d_2}{2} \, \widetilde{F}_{B\mu\nu} (\widetilde{F}_{B\kappa\lambda}f^{\kappa\lambda})
%-\frac{3}{4} \, R_{0\mu\nu\kappa\lambda\rho\sigma}f^{\kappa\lambda}f^{\rho\sigma}
%\right.
%\nonumber \\
%\left.
%-\frac{1}{2} \, S_{0\mu\nu\kappa\lambda\rho\sigma\omega\tau}f^{\kappa\lambda}f^{\rho\sigma}f^{\omega\tau} \,
\right]=0,
%-\,c_{1}\,\partial^{\mu}F_{0\mu\nu}
\end{equation}
and the dual tensor $\widetilde{f}^{\mu\nu}$ satisfies the Bianchi identity $\partial_{\mu}\widetilde{f}^{\mu\nu}=0$.
For the analysis of the propagation effects, we consider the Lagrangian density (\ref{L2}).
The correspondent field equation  are similar to eqs. (\ref{eqdivD}) and (\ref{eqdivB}), just exchanging
$({\bf E},{\bf B})$ by $({\bf e},{\bf b})$, where ${\bf D}=\epsilon_{ij}({\bf B}_{0}) \, e_{j}$ and ${\bf H}=(\mu_{ij})^{-1}({\bf B}_{0}) \, b_{j}$,
in which the permittivity and the permeability tensors are functions of the magnetic background components :
\begin{subequations}
\begin{eqnarray}
\epsilon_{ij}({\bf B}_{0}) \!&=&\! \delta_{ij} + \frac{d_2}{c_{1}} \, B_{0i} \, B_{0j} \; ,
\label{epsilonijB}
\\
\mu_{ij}({\bf B}_{0}) \!&=&\! \delta_{ij}+\frac{d_1 \, B_{0i}\,B_{0j}  }{c_1-d_1\,{\bf B}_{0}^{2}} \; .
\label{muijB}
\end{eqnarray}
\end{subequations}
The eigenvalues of the permittivity and permeability matrices are given by
\begin{subequations}
\begin{eqnarray}
\lambda_{1\varepsilon} \!&=&\! \lambda_{2\varepsilon} = 1
\; , \;
\lambda_{3\varepsilon}({\bf B}_{0}) = 1 + \frac{d_{2}}{c_{1}} \, {\bf B}_{0}^2 \; ,
\\
\lambda_{1\mu} \!&=&\! \lambda_{2\mu} = 1
\; , \;
\lambda_{3\mu}({\bf B}_{0}) = \frac{c_{1}}{c_{1}-d_1\,{\bf B}_{0}^2}  \; .
\end{eqnarray}
\end{subequations}
In the limit $\beta \rightarrow \infty$, the matrices (\ref{epsilonijB}) and (\ref{muijB}) reduce to identity matrix,
and the eigenvalues go to one. Note that these limits are also recovered by turning off the magnetic background field.
The permittivity matrix is positive for any background magnitude, while the permeability matrix is positive if
we impose the conditions $c_{1}>0$ and $c_1>d_{1}\,{\bf B}_{0}^2$. Under these conditions, the material medium behaves
like a paramagnetic material. Using the relations $\epsilon_{ij}=\delta_{ij}+\chi_{Eij}$ and $\mu_{ij}=\delta_{ij}+\chi_{Mij}$,
the electric and magnetic susceptibility tensors are, respectively, read below :
\begin{subequations}
\begin{eqnarray}
\chi_{Eij} ({\bf B}_{0}) \!&=&\! \frac{d_2}{c_{1}} \, B_{0i} \, B_{0j} \; ,
\label{chiEij}
\\
\chi_{Mij} ({\bf B}_{0}) \!&=&\! \frac{d_1 \, B_{0i}\,B_{0j}  }{c_1-d_1\,{\bf B}_{0}^{2}} \; .
\label{chiMij}
\end{eqnarray}
\end{subequations}
In addition, the eigenvalues of the susceptibility matrices are given by
\begin{subequations}
\begin{eqnarray}
\chi_{1E} \!&=&\! \chi_{2E} = 0
\; , \;
\chi_{3E}({\bf B}_{0}) = \frac{d_{2}}{c_{1}} \, {\bf B}_{0}^2 \; ,
\\
\chi_{1M} \!&=&\! \chi_{2M} =0
\; , \;
\chi_{3M}({\bf B}_{0}) = \frac{1}{c_{1}-d_1\,{\bf B}_{0}^2}  \; ,
\end{eqnarray}
\end{subequations}
where the magnetic susceptibility is positive if we also impose the condition $c_{1}>d_1\,{\bf B}_{0}^2$.
Contracting the field eq. (\ref{eqf}) with $f_{\nu\alpha}$ and using the Bianchi identity, we obtain
the conservation law
\begin{eqnarray}\label{eqTheta}
\partial_{\mu}\Theta^{\mu}_{\;\;\alpha}=0 \; ,
\end{eqnarray}
with the following energy-momentum tensor
\begin{eqnarray}
\Theta^{\mu}_{\;\;\,\alpha} \!&=&\! c_{1} \, f^{\mu\nu}f_{\nu\alpha}
-\frac{d_1}{2}\, (F_{B}^{\;\,\,\mu\nu}f_{\nu\alpha})(F_{B\kappa\lambda}f^{\kappa\lambda})
\nonumber \\
&&
\hspace{-0.5cm}
-\frac{d_2}{2}\, (\widetilde{F}_{B}^{\;\,\,\mu\nu}f_{\nu\alpha})(\widetilde{F}_{B\kappa\lambda}f^{\kappa\lambda})
-\delta^{\mu}_{\;\;\,\alpha} \, {\cal L}^{(2)}
%-\frac{3}{4}\,R_{0}^{\;\,\mu\nu\kappa\lambda\rho\sigma}f_{\kappa\lambda}f_{\rho\sigma}f_{\nu\alpha}
%\nonumber \\
%&&
%\hspace{-0.5cm}
%-\frac{1}{2}\,S_{0}^{\;\,\mu\nu\kappa\lambda\rho\sigma\omega\tau}f_{\kappa\lambda}f_{\rho\sigma}f_{\omega\tau}f_{\nu\alpha}
\; .
\end{eqnarray}
Notice that the eq. (\ref{eqTheta}) is due to the uniform magnetic background field. The conserved
components of the energy-momentum tensor are $\Theta^{\mu0}=\left(\Theta^{00},\Theta^{i0}\right)$ in which
the energy density and the Poynting vector components can be written as
\begin{subequations}
\begin{eqnarray}
\Theta^{00} \!&=&\! \frac{1}{2} \, \epsilon_{ij}({\bf B}_{0}) \, e_{i} \, e_{j}
+\frac{1}{2} \, (\mu_{ij})^{-1}({\bf B}_{0}) \, b_{i} \, b_{j} \; ,
\\
\Theta^{i0} \!&=&\!  c_1 \, ({\bf e} \times {\bf b} )^i + d_1 \, ({\bf e} \cdot {\bf E})\;({\bf e} \times {\bf B} )^i
\nonumber \\
&&
\hspace{-0.5cm}
-\,d_1 \, ({\bf b} \cdot {\bf B})\;({\bf e} \times {\bf B} )^i
-\,d_2 \, ({\bf e} \cdot {\bf B})\;({\bf e} \times {\bf E} )^i
\nonumber \\
&&
\hspace{-0.5cm}
-d_2 \, ({\bf b} \cdot {\bf E})\;({\bf e} \times {\bf E} )^i \, .
\end{eqnarray}
\end{subequations}
Thereby, the energy density is so positive if it also satisfies the condition $c_1>d_{1}{\bf B}_{0}^2$.
The plane wave solutions for the propagating fields ${\bf e}$ and ${\bf b}$
are given in the usual form :
\begin{eqnarray}
{\bf e}({\bf r},t)={\bf e}_{0} \, e^{i({\bf k}\cdot{\bf r}-\omega t)}
\;\,\, \mbox{and} \;\,\,
{\bf b}({\bf r},t)={\bf b}_{0} \, e^{i({\bf k}\cdot{\bf r}-\omega t)} \; ,
\end{eqnarray}
in which the electric ${\bf e}_{0}$ and magnetic ${\bf b}_{0}$ amplitudes are
related by the Faraday's law as ${\bf b}_{0}=({\bf k}\times{\bf e}_{0})/\omega$,
where ${\bf k}$ denotes the wave vector, and $\omega$ is the frequency. Substituting these
solutions in the field equations, the wave equation of the electric amplitude is
\begin{eqnarray}
M_{ij}\,e_{0j}=0 \; ,
\end{eqnarray}
in which the matrix elements of $M_{ij}$ are written as
\begin{eqnarray}
M_{ij} \!&=&\! \left[ 1-{\bf n}^2+\frac{d_1}{c_1} \left({\bf n} \times {\bf B}_{0}\right)^2 \right] \delta_{ij} +
\nonumber \\
&&
\hspace{-0.6cm}
+ \left(1-\frac{d_1}{c_1} \, {\bf B}_{0}^2\right) n_i \, n_j
+\left( \frac{d_2}{c_1} - \frac{d_1}{c_1} \, {\bf n}^2 \right) B_{0i} \, B_{0j}
%\nonumber \\
%&&
%\hspace{-1cm}
\nonumber \\
&&
\hspace{-0.6cm}
+\frac{d_1}{c_1} \left({\bf B}_{0} \cdot {\bf n}\right) \left(B_{0i} \, n_j + B_{0j} \, n_i\right)
%
%\nonumber \\
%&&
%\hspace{-1cm}
%+\frac{g^2}{c_1} \, \frac{[ \, \omega B_{i} - ({\bf k}\times{\bf E})_{i} \, ][ \, \omega B_{j} - ({\bf k}\times{\bf E})_{j} \, ] }{{\bf k}^2-\omega^2+m^2}
\; ,
\end{eqnarray}
where we have defined ${\bf n}={\bf k}/\omega$. The null determinant $\mbox{det}(M)=0$ yields the ${\bf n}$-polynomial equation
\begin{eqnarray}
\left[1-{\bf n}^2+\frac{d_1}{c_1}\,({\bf n}\times{\bf B}_{0})^{2}\right] \times
\nonumber \\
\times\left[ \left(1+\frac{d_2}{c_1}\,{\bf B}_{0}^2 \right)(1-{\bf n}^2)+\frac{d_2}{c_1}\,({\bf n}\times{\bf B}_{0})^2 \right]=0 \; ,
\end{eqnarray}
whose solutions provide two possible refractive indices
\begin{subequations}
\begin{eqnarray}
n_1 \!&=&\! \sqrt{\frac{c_1}{c_1-d_1\,({\bf B}_{0}\times\hat{{\bf k}})^2}} \; ,
\\
n_2 \!&=&\! \sqrt{ \frac{c_1+d_2\,{\bf B}_{0}^2}{c_1+d_2\,({\bf B}_{0}\cdot\hat{{\bf k}})^{2}} } \; .
\end{eqnarray}
\end{subequations}
Using the coefficients from eq. (\ref{coeficientsresult}), the refractive indices are functions of the angle $\theta$
between the magnetic background field and wave propagation direction $\hat{{\bf k}}$. In the limit
$\beta\rightarrow\infty$, both the refractive indices go to one. Moreover, in the regime of a strong magnetic field
$(|{\bf B}_{0}|\gg\beta)$, the first solution is $n_{1}\simeq 1$, and the second refractive index depends only
on the $\theta$-angle, namely,
\begin{eqnarray}
n_{2}\simeq |\sec\theta| \; .
\end{eqnarray}
For the TaAs Dirac material listed in the table (\ref{table1}), $\beta \rightarrow 0$,
and the refractive index have the same results, $n_{1}=1$ and $n_{2}=|\sec\theta|$, respectively.
In the case of the QED, we consider $\alpha_{D}=\alpha = 0.0073$, and in this approximation,
the refractive index are given by
\begin{subequations}
\begin{eqnarray}
n_1 \!&\simeq&\! 1+\frac{\alpha}{24\pi^2}\frac{{\bf B}_{0}^2+2\beta^2}{({\bf B}_{0}^2+\beta^2)^{2}}
\,(\hat{{\bf k}}\times{\bf B}_{0})^{2}
%+{\cal O}(\alpha_{D}^2)
\; ,
\\
n_2 \!&\simeq&\! 1+\frac{\alpha}{24\pi^2}\frac{|{\bf B}_{0}|}{\beta}
\,\frac{(\hat{{\bf k}}\times{\bf B}_{0})^{2}}{{\bf B}_{0}^2+\beta^2}
%+{\cal O}(\alpha_{D}^2)
\; .
\end{eqnarray}
\end{subequations}
In the regime of a strong magnetic field, the refraction index in the QED are reduced to the results :
\begin{subequations}
\begin{eqnarray}
n_1 \!&\simeq&\! 1+\frac{\alpha}{24\pi^2} \, \sin^2\theta
%+{\cal O}(\alpha_{D}^2)
\; ,
\\
n_2 \!&\simeq&\! 1+\frac{\alpha}{24\pi^2}\frac{|{\bf B}_{0}|}{\beta}
\, \sin^2\theta
%+{\cal O}(\alpha_{D}^2)
\; .
\end{eqnarray}
\end{subequations}
The first solution does not depend on the magnitude of the magnetic background, and it is
function of the $\theta$-angle. This conclusion shows that the correspondent dispersion
relations does not depend on the magnitude of the magnetic background field in the
regime of a strong background, and it depends only on the $\theta$-angle.
\section{Birefringence phenomenon in a uniform magnetic background}
\label{sec4}
\begin{figure}[th]
%\vspace{-5pt}
\centering
\includegraphics[width=0.48\textwidth]{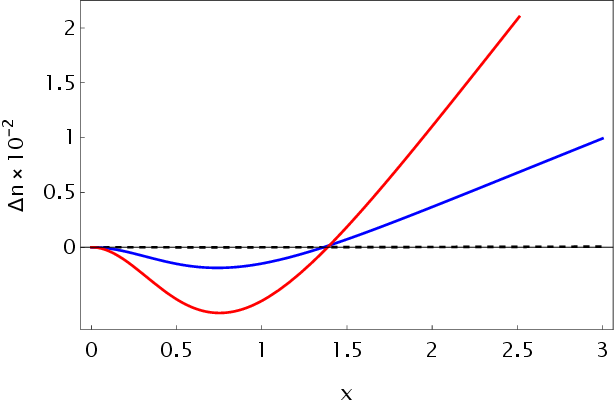}
%\quad\quad
%\includegraphics[width=0.45\textwidth]{n2x.eps}
\caption{The variation of the refractive index $(\Delta n)$ versus the dimensionless variable $x=B_{0}/\beta$
for the Dirac materials BiSb (blue line) and PbSnTe (red line).}
\label{deltanB}
\end{figure}

We now consider the birefringence phenomenon, which rises from the wave propagation associated with different amplitudes. In this case,
we consider the magnetic background on the ${z}$-direction, {\it i.e.}, ${\bf B}_{0}=B_{0}\,\hat{{\bf z}}$,
and in the first situation, the plane wave solution for the electric field has an amplitude parallel
to the magnetic background, where ${\bf e}({\bf r},t)=e_{03}\,\hat{{\bf z}} \, e^{i({\bf k}\cdot{\bf r}-\omega t)}$.
Using this solution, the wave equation can be written as
\begin{eqnarray}
\left[ \, \frac{{\bf k}^2}{\omega^2}- \epsilon_{33}(B_{0}) \, \mu_{22}(B_{0}) \, \right]e_{03}=0 \; ,
\end{eqnarray}
in which the correspondent refractive index is
\begin{eqnarray}\label{nparalelB}
n_{\parallel}(B_{0})=\sqrt{ \mu_{22}(B_{0}) \, \epsilon_{33}(B_{0}) } = \sqrt{ \, 1+ \frac{d_2}{c_1}\,B_{0}^2 \, } \; .
\end{eqnarray}
The second situation happens when the plane wave has amplitude perpendicular to the magnetic background, such that
${\bf e}({\bf r},t)=e_{02}\,\hat{{\bf y}} \, e^{i({\bf k}\cdot{\bf r}-\omega t)}$. In this case, the wave equation is
\begin{eqnarray}
\left[ \, \frac{{\bf k}^2}{\omega^2}- \epsilon_{22}(B_{0})\,\mu_{33}(B_{0}) \, \right]e_{02}=0 \; ,
\end{eqnarray}
with the refractive index
\begin{eqnarray}
n_{\perp}(B_{0})=\sqrt{ \mu_{33}(B_{0})\,\epsilon_{22}(B_{0}) } = \sqrt{ \, \frac{c_{1}}{c_{1} - d_{1} \, B_{0}^2} \, } \; .
\end{eqnarray}
The birefringence phenomena rises from the difference
\begin{eqnarray}\label{Deltan}
\Delta n(B_{0}) = n_{\parallel}(B_{0})-n_{\perp}(B_{0}) \neq 0 \; .
\end{eqnarray}
Therefore, the birefringence does not occur if we constraint the condition
\begin{eqnarray}
\mu_{22}(B_{0})\,\epsilon_{33}(B_{0})=\epsilon_{22}(B_{0})\,\mu_{33}(B_{0}) \; ,
\end{eqnarray}
and using the $22$ and $33$ components from eqs. (\ref{epsilonijB}) and (\ref{muijB}), we obtain the relation :
\begin{eqnarray}\label{d1d2}
d_{2}-d_{1}=\frac{d_1\,d_2}{c_1} \, B_{0}^2 \; .
\end{eqnarray}
In the limit $\beta \rightarrow \infty$, the birefringence disappears in eq. (\ref{Deltan}).
For $\beta \gg B_{0}$, this difference is residual in $(B_{0}/\beta)^2$,
\begin{eqnarray}\label{Deltanapprox}
\Delta n(B_{0}) \simeq - \frac{\alpha_{D}}{12\pi^2} \left(\frac{B_{0}}{\beta}\right)^2 \; .
\end{eqnarray}
The plot of $\Delta n$ versus the dimensionless variable $x=B_{0}/\beta$ is illustrated in the fig. (\ref{deltanB}),
for the Dirac materials from the table \ref{table1} : BiSb (blue line) and PbSnTe (red line).
The lines intercept the horizontal axis where the birefringence is null in accord with the condition (\ref{d1d2}). Using the coefficients (\ref{coeficientsresult}) and the physical parameters in the table \ref{table1}, we obtain the magnitude of the magnetic background field
for the QED and the Dirac materials in which the birefringence is null :
\begin{subequations}
\begin{eqnarray}
B_{0} \!\!&=&\!\! 4.0 \; \mbox{MeV}^2=5.6 \, \mbox{GT} \hspace{0.2cm} \mbox{for} \hspace{0.2cm} \mbox{QED} \; ,
\label{B0resultQED}
\\
B_{0} \!\!&=&\!\! 8.90 \; \mbox{eV}^2 = 12 \, \mbox{mT} \hspace{0.2cm} \mbox{for} \hspace{0.2cm} \mbox{PbSnTe} \; ,
\label{B0resultPBSnTe}
\\
B_{0} \!\!&=&\!\! 0.18 \; \mbox{eV}^2 = 0.25 \, \mbox{mT} \hspace{0.2cm} \mbox{for} \hspace{0.2cm} \mbox{BiSb} \; .
\label{B0resultBiSb}
\end{eqnarray}
\end{subequations}
Otherwise, the condition (\ref{Deltan}) is kept for values of $B_{0}$ different from the results
(\ref{B0resultQED})-(\ref{B0resultBiSb}).
\section{Interaction energy for an electrodynamics in Dirac materials under a uniform magnetic field}
\label{sec5}
As already expressed, here we calculate the interaction energy between static point-like sources for an electrodynamics in Dirac materials under a uniform magnetic field, along the lines of refs. \cite{Gaete97,GEN_BI,LOG,Gaete_AHEP_2021,Gaete_EPJC_2022}. To do that, we compute the expectation value of the energy operator $H$ in the physical state $\left| \Phi  \right\rangle$, which we write as $\langle H\rangle _\Phi$.
In this perspective, our aim here is to evaluate to the lowest-order the interaction energy. For this purpose, we linearize the effective theory (\ref{Lmodel}) in a similar manner to that which led to the eq. (\ref{L2}). Furthermore, in what follows, we confine our attention to the case of a
pure and uniform magnetic background (${\bf E}={\bf 0}$), as in the section \ref{sec3}.
In such a case, we back to the effective Lagrangian density in eq. (\ref{L2}).
%reads
%
%\begin{equation}
%{\cal L} =  - \frac{1}{4}{c_1}{f^2} + \frac{{{d_1}}}{8}{\left( {{F_{B\mu \nu }}{f^{\mu \nu }}} \right)^2} + \frac{{{d_2}}}{8}{\left( {{{\tilde F}_{B\mu \nu }}{f^{\mu \nu }}} \right)^2}.
%\label{intdmat05}
%\end{equation}
%

%
We now carry out a Hamiltonian analysis of this theory. The canonical momenta following from eq. (\ref{L2}) are
\begin{equation}
{\Pi ^\mu } =  - {c_1}{f^{0\mu }} + \frac{{{d_1}}}{2}F_{B}^{\;\,\,\rho \sigma }{f_{\rho \sigma }}F_{B}^{\;\,\,0\mu }
+ \frac{{{d_2}}}{2} \, \tilde F_{B}^{\;\,\,\rho \sigma } f_{\rho \sigma } \, \tilde F_{B}^{\;\,\,0\mu }, \label{intdmat10}
\end{equation}
which produces the usual primary constraint $\Pi^0 = 0$, and
%\begin{equation}
${\Pi _i} = {c_1}\left( {{\delta _{ij}} + \frac{{{d_2}}}{{{c_1}}}\,{B_{0i}}\,{B_{0j}}} \right){e_j}$ \; .
%\label{intdmat15-b}
%\end{equation}

Making use of this last equation, we readily verify that the electric field associated to the  $a^{\mu}$-field takes the form
\begin{equation}
{e_i} = \frac{1}{{{c_1}\det D}}\left( {{\delta _{ij}}\det D - \frac{{{d_2}}}{{{c_1}}} \, {B_{0i}}{B_{0j}}} \right){\Pi _j} \, .
\label{intdmat20}
\end{equation}
Here $\det D = 1 + \frac{{{d_2}}}{{{c_1}}}\,{{\bf B}_{0}^2}$, where ${\bf B}_{0}$ represents the external (background) magnetic field around which the $a^{\mu}$-field fluctuates.
The canonical Hamiltonian is now obtained in the usual way by a Legendre transformation to obtain
\begin{eqnarray}
{H_C} &=& \int {{d^3}x} \left[ \, {{\Pi _i}{\partial ^i}{a_0} + \frac{{{{\bf \Pi} ^2}}}{{2\,{c_1}}} + \frac{{{c_1}}}{2}\,{{\bf b}^2} - \frac{{{d_1}}}{2}\,{{\left( {{\bf B}_{0} \cdot {\bf b}} \right)}^2}} \, \right] \nonumber\\
&-& \int {{d^3}x} \, \frac{{{d_2}}}{{2\,c_1^2\det D}} \, {\left( {{\bf B}_{0} \cdot {\bf \Pi} } \right)^2}.
\label{intdmat25}
\end{eqnarray}
It can be quickly checked that the preservation in time of the primary constraint, ${\Pi}_0$, leads to the secondary constraint ${\Gamma _1} \equiv {\partial _i}{\Pi ^i} = 0$ (Gauss constraint). In this case, there are two constraints, which are first-class ones. According to the general theory, we obtain the extended Hamiltonian, which generates translations in time, by adding all the first-class constraints with arbitrary coefficients to the canonical Hamiltonian. Accordingly, we write
\begin{eqnarray}
H = H_C + \int {d^3 } x\left[ \, {c_0 \left( x \right) \Pi _0 \left( x \right) + c_1
\left( x\right)\Gamma _1 \left( x \right)} \, \right], \label{intdmat30}
\end{eqnarray}
where $c_0 \left( x\right)$ and $c_1 \left( x \right)$ are arbitrary functions of the space-time coordinates. Since  $\Pi^0=0$ always, neither $a^0 $ nor $\Pi^0$ are of interest in describing the system and may be excluded from the theory. Thus the variables $a^0 $ and $\Pi^0$  play no role in the theory. If a new arbitrary coefficient $c(x) = c_1 (x) - a_0 (x)$ is introduced the Hamiltonian may be rewritten in the form
\begin{eqnarray}
H &=& \int {{d^3}x} \left[ \, {c\left( x \right){\partial ^i}{\Pi _i} + \frac{{{{\bf \Pi} ^2}}}{{2\,{c_1}}} + \frac{{{c_1}}}{2}\,{{\bf b}^2} - \frac{{{d_1}}}{2}{{\left( {{\bf B}_{0} \cdot {\bf b}} \right)}^2}} \, \right]
\nonumber \\
 &-& \int {{d^3}x} \, \frac{{{d_2}}}{{2\,c_1^2\det D}} \, {\left( {{\bf B}_{0} \cdot {\bf \Pi} } \right)^2}.
 \label{intdmat35}
\end{eqnarray}
Since there is one first constraint, ${\Gamma _1}$, we choose a gauge-fixing condition that will make the full set of constraints to become second class. To do this, we adopt the gauge discussed previously \cite{Gaete97}, that is,
\begin{equation}
\Gamma _2 \left( x \right) \equiv \int\limits_{C_{\zeta x} } {dz^\nu }
\, a_\nu\left( z \right) \equiv \int\limits_0^1 {d\lambda \, x^i } a_i \left( {
\lambda x } \right) = 0.  \label{intdmat40}
\end{equation}
Here $\lambda$ $(0\leq \lambda\leq1)$ is the parameter describing the
space-like straight path $x^i = \zeta ^i + \lambda \left( {x - \zeta}
\right)^i $, and $\zeta^{i}$ is a fixed point (reference point).
There is no essential loss of generality if we restrict our considerations to $\zeta^i=0$.
With such a choice, the fundamental Dirac bracket is given by
\begin{eqnarray}
\left\{ {a_i \left( {\bf x} \right),\Pi ^j \left( {\bf y} \right)} \right\}^{\ast} \!&=&\! \delta_{i}^{\;\,j} \, \delta ^{\left( 3 \right)} \left( {{\bf x} - {\bf y}} \right)
\nonumber \\
&&
\hspace{-0.5cm}
-\partial_i^x
\int\limits_0^1 {d\lambda \, x^j } \delta ^{\left( 3 \right)} \left( {\lambda
{\bf x}- {\bf y}} \right).   \label{intdmat45}
\end{eqnarray}
We now recall that the physical states $\left| \Phi \right\rangle $ are gauge-invariant. In this case, we may therefore write the stringy gauge-invariant state
\begin{eqnarray}
\left| \Phi  \right\rangle  \!&\equiv&\! \left|\, {\overline{\Psi} \left( {\bf y} \right)\Psi \left( {{{\bf y}^ {\prime} }} \right)} \,\right\rangle  \nonumber\\
&=& \overline{\Psi} \left( {\bf y} \right)\exp \left[ {iq\int_{{{\bf y}^ {\prime} }}^{\bf y} {d{z^i}{a_i}\left( z \right)} } \right]\Psi \left( {{{\bf y}^ {\prime} }} \right)\left| 0 \right\rangle,  \label{intdmat50}
\end{eqnarray}
where the line integral is along a space-like path on a fixed time slice, $q$ is the fermion charge and $\left| 0 \right\rangle$ is the physical vacuum state.

\begin{figure}[th]
%\vspace{-5pt}
\centering
\includegraphics[width=0.48\textwidth]{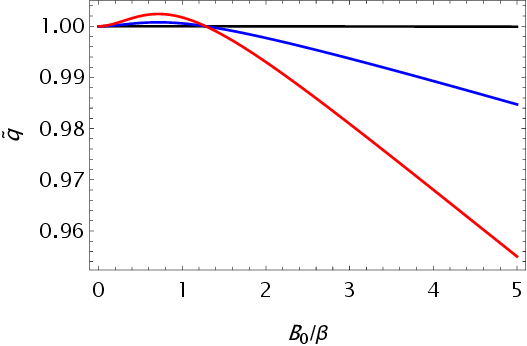}
%\quad\quad
%\includegraphics[width=0.45\textwidth]{n2x.eps}
\caption{The dimensionless effective charge $\tilde{q}$ as function of the ratio $B_{0}/\beta$ (magnetic background field over critical field)
for the QED (black line), and for the Dirac materials BiSb (blue line) and PbSnTe (red line).
%for the Dirac materials BiSb (blue line) and PbSnTe (red line).
}
\label{qeff}
\end{figure}

At this point, we have all the elements necessary for the calculation of the expectation value ${\left\langle H \right\rangle _\Phi }$. From the foregoing discussion we readily see that ${\left\langle H \right\rangle _\Phi }$ can therefore be written as follows
\begin{equation}
{\left\langle H \right\rangle _\Phi } = {\left\langle H \right\rangle _0} + \left\langle H \right\rangle _\Phi ^{\left( 1 \right)}, \label{intdmat55}
\end{equation}
where ${\left\langle H \right\rangle _0} = \left\langle 0 \right|H\left| 0 \right\rangle$, whereas the $\left\langle H \right\rangle _0^{\left( 1 \right)}$ term is given by
\begin{equation}
\left\langle H \right\rangle _\Phi ^{\left( 1 \right)} = \left\langle  \Phi  \right|\int {{d^3}x} \left[ {\frac{{{{\bf \Pi} ^2}}}{{2\,{c_1}}} -\frac{{{d_2}}}{{2\,c_1^2\det D}}{{\left( {{\bf B}_{0} \cdot {\bf \Pi} } \right)}^2}} \right]\left| \Phi  \right\rangle, \label{intdmat60}
\end{equation}
Following our earlier procedure \cite{Gaete_AHEP_2021,Gaete_EPJC_2022}, the static potential profile for two opposite charges located at ${\bf y}$ and ${\bf y}^{\prime}$ then reads
\begin{equation}\label{intdmat65}
V(L) = - \frac{{{q_{eff}^2({\bf B}_{0})}}}{{4\pi }}\frac{1}{L} \; ,
\end{equation}
where the effective charge is
\begin{eqnarray}
q_{eff}({\bf B}_0) = \frac{|q|}{\sqrt{c_1 + d_2 \, {{\bf B}_{0}^2}}} \; ,
\end{eqnarray}
and $L \equiv |{\bf y} - {{\bf y}^ {\prime} }|$ is the distance that separates the two charges. We see, therefore, that at the lowest-order in our effective model, the introduction of the non-linearities induce redefinition in the charges. Whenever $|{\bf B}_{0}| \rightarrow 0$, the effective charge $q_{eff}$ reduces to the usual $q$-charge. This result agrees with the absence of screening in the Coulomb potential that is a characteristic of Dirac materials. Under an intense magnetic background, $q_{eff}$ goes to zero, and consequently, the potential (\ref{intdmat65}) signals a situation of free particles. The dimensionless quantity $\tilde{q}=q_{eff}/|q|$ is plotted as function of $B_{0}/\beta$ in fig. (\ref{qeff}), for the QED case (black line)
and the Dirac materials BiSb (blue line) and PbSnTe (red line).

\noindent
The QED curve is approximately horizontal since the contribution of the fine structure
constant is very weak in this case. When the fine structure constant increases
(that are the cases of the BiSb and PbSnTe), there arises a peak showed in the blue and red lines, respectively.
In the case of the PbSnTe, the peak is localized at magnetic background $B_{0}=4.70 \, \mbox{eV}^2$,
where the effective charge is maximum at $q_{eff}=0.3035$ for two electrons interacting under the potential (\ref{intdmat65}).

\section{Concluding comments and considerations}
\label{sec6}
In this contribution, a non-linear electrodynamics (ED) similar to a non-linear logarithmic model is proposed as a possible formulation from which there follows the effective Lagrangian density in ref.
\cite{Keser}, obtained from the one-loop radiative corrections for Dirac materials. We highlight that, in the regime of a strong magnetic field related to electric field, the non-linear ED under consideration reduces to the situation presented in ref. \cite{Keser}. The non-linear model introduces a critical electric field as a parameter that depends on the characteristics of the particular Dirac material. The usual Maxwell electrodynamics is recovered in the limit in which the critical electric field ($\beta$-parameter) goes to infinity. Using the non-linear Lagrangian density, we obtain the non-linear field equations, the permittivity and permeability tensors, and the correspondent energy-momentum tensor.

Afterwards, we study the aspects of the model when the system is submitted to an external (and uniform) magnetic
field. We expand the non-linear Lagrangian up to second order in the propa\-gating electromagnetic field. Thereby, the permittivity and permeability tensors are obtained in terms of the magnetic background field components. The refractive indices are calculated as function of the $\beta$-parameter and magnetic background field. We observe that it also depends on the angle $(\theta)$ in that the propagation direction does with magnetic background field. In the regime of a strong magnetic background, one of the solutions for the refractive index depends only on the $\theta$-angle. Furthermore, we investigate the birefringence phenomenon for the wave electric field propagating in this non-linear theory under a uniform magnetic background. We calculate numerically the magnetic background values for which birefringence does not take place; to do that, we must use physical characteristics of the particular Dirac material.

In the paper of Ref. \cite{Mola}, the authors inspect the interesting question of the propagation of an undamped transverse electric mode due to charge density fluctuations in a type of material known as tilted Dirac-cone material  (TDM). Charge-density excitations referred to as plasmons are actually present in Dirac semimetals; in the work cited as Ref. \cite{Kharzeev}, the authors discuss possible practical applications of plasmon excitations with frequency range in the THz region of the spectrum. Stimulated by the results of the references quoted above and by the interest in studying surface plasmons in Dirac semimetals, we have in project to re-assess these issues by applying our developments based on the Lagrangian density of eq. (\ref{LDM}) to specific Dirac-type materials. The permittivity tensor presented in eq. (\ref{epsilonij}) plays a key role in our endeavor.

In Section IV, we devote efforts to investigate the phenomenon of birefringence associated to the model described by the Lagrangian density of eq. (4). By considering the graph depicted in Fig. 2, we may propose, as a possible experimental observation, the measurement of the birefringence as given by $\Delta n(B_0)$, according to eq. (\ref{Deltan}). As the graph illustrates, for laboratory magnetic fields, corresponding to values of $x$ between $2.0$ and $2.5$, for example, $\Delta n$ presents accessible values for the BiSb and PbSnTe materials. This could provide a way of testing our results on birefringence. 

To conclude, we apply the path-dependent formalism to the linearized Lagrangian to obtain the interaction energy between two static point-like charges. There results a Coulomb-like potential with an effective charge that runs with the magnetic background field. For the PbSnTe Dirac material, the effective charge has a smooth peak at a magnetic field of $4.70 \, \mbox{eV}^2$, as illustrated in fig. \ref{qeff}.
Taking a forward-looking perspective, as a possible path to further exploit the effective photonic model in discussion here, we believe that a natural follow-up would be the inspection of the regime of vortex formation and the consequent study of the properties of the magnetic vortices governed by the set of the non-linear electromagnetic field equations. As a consequence of the non-linearity, charged vortices might naturally arise and investigating the dynamics of fermions in presence of such vortices may be a research question of interest in the context of Dirac materials.
\section{Acknowledgments}
L.P.R. Ospedal is grateful to FAPERJ (grant number E-26/203.997/2022) for his post-doctoral fellowship.
P. Gaete was partially supported by ANID PIA/APOYO AFB220004.

%ANID PIA / APOYO AFB180002.

\hspace{0.5cm}

%{\bf Data Availability Statement: No Data associated in the manuscript.}


\begin{thebibliography}{30}

\bibitem{Nagaosa} N. Nagaosa, \textit{Quantum Field Theory in Condensed Matter} Physics, Springer, 1999.

\bibitem{Tsvelik} A. M. Tsvelik, \textit{Quantum Field Theory in Condensed Matter Physics}, Cambridge, 2003.

\bibitem{Keser} Aydin C. Keser, Yuli Lyanda-Geller and Oleg P. Sushkov, Phys. Rev. Lett. {\bf 128}, 066402, 2022.

\bibitem{Euler} H. Euler and W. Heisenberg, Z. Phys. {\bf 98}, 714 (1936).

\bibitem{Schwinger} J. Schwinger, Phys. Rev. {\bf 82}, 664 (1951).

\bibitem{Adler} S.L. Adler, Ann. Phys. (N.Y.) {\bf 67}, 599 (1971).

\bibitem{Costantini} V. Costantini, B. De Tollis, G. Pistoni, Nuovo Cimento A {\bf 2}, 733 (1971).

\bibitem{Ruffini} R. Ruffini, G. Vereshchagin, S.-S. Xue, Phys. Rep. {\bf 487}, 1-140
(2010).

\bibitem{Dunne} G.V. Dunne, Int. J. Mod. Phys. Conf. Ser. {\bf 14}, 42 (2012).

\bibitem{Battesti} R.Battesti, C.Rizzo, Rep.Prog.Phys. {\bf 76}, 016401 (2013).

\bibitem{Sarazin} X. Sarazin, F. Couchot, A. Djannati-Ata\"{\i}, O. Guilbaud, S. Kazamias, M. Pittman, M. Urban, Eur. Phys. J. D {\bf 70}, 13 (2016).

\bibitem{ATLAS} ATLAS Collaboration, Nat. Physical. {\bf 13}, 852 (2017).

\bibitem{CMS} CMS Collaboration, Phys. Lett. B {\bf 797}, 134826 (2019).

\bibitem{Schoeffel} L. Schoeffel, C. Baldenegro, H. Hamdaoui, S. Hassani, C. Royon and M. Saimpert, Prog. Part. Nucl. Phys. {\bf 120}, 103889 (2021).

\bibitem{Robertson2} S. Robertson, A. Mailliet, X. Sarazin, F. Couchot, E. Baynard, J. Demailly,  M. Pittman, A. Djannati-Ata\"\i{}, S. Kazamias, and M. Urban, Phys. Rev. A {\bf 103}, 023524 (2021).

\bibitem{Battesti2}  R. Battesti {\it et al}, Phys. Rept. {\bf 765-766}, 1-39, (2018).

\bibitem{Ataman} S. Ataman, Phys. Rev. A  {\bf 97}, 063811 (2018).

\bibitem{Wehling} T. O. Wehling, A. M. Black-Schaffer and A. V. Balatsky, {\it Dirac materials}, Adv. Phys. {\bf 76} (2014) 1.

\bibitem{MJNevesEDN} M. J. Neves, Jorge B. de Oliveira, L. P. R. Ospedal and J. A. Helay\"el-Neto,  Phys. Rev. D {\bf 104} (2021) 015006.

\bibitem{Gaete97} P. Gaete, Z. Phys. C, {\bf 76} 355 (1997).

\bibitem{GEN_BI} P.~Gaete and J.~Helay\"el-Neto, Eur.\ Phys.\ J.\ C {\bf 74}, 3182 (2014).

\bibitem{LOG} P.~Gaete and J.~A.~Helay\"el-Neto, Eur. Phys. J. C \textbf{81}, 899 (2021).

\bibitem{Gaete_AHEP_2021} P. Gaete, J.A. Helay\"el-Neto and L.P.R. Ospedal, Adv. High Energy Phys. {\bf 2021}, 6621975 (2021).

\bibitem{Gaete_EPJC_2022} M.~J.~Neves, L.~P.~R.~Ospedal, J.~A.~Helay\"el-Neto and P.~Gaete, Eur. Phys. J. C \textbf{82}, no.4, 327 (2022).


\bibitem{Mola} Z. Jalali-Mola and S. A. Jafari,
%Undamped transverse electric mode in undoped two-dimensional tilted Dirac cone materials,
Phys. Rev. B {\bf 102} (2020) 245148.


\bibitem{Kharzeev} D. E. Kharzeev, R. D. Pisarski and H.-U. Yee,
%Universality of plasmon excitations in Dirac semimetals,
Phys. Rev. Lett. {\bf 115} (2015) 236402.


%\bibitem{GaeteEPJC1} Patricio Gaete and José Helayël-Neto, {\it Finite field-energy and interparticle potential in logarithmic electrodynamics}, Eur. Phys. J. C (2014) {\bf 74} 2816.

%
%\bibitem{GaeteEPJC2} Patricio Gaete and José Helayël-Neto, {\it Remarks on non-linear electrodynamics}, Eur. Phys. J. C (2014) {\bf 74} 3182.

%
%\bibitem{Silva} Pedro D. S. Silva, Rodolfo Casana and Manoel M. Ferreira Jr., {\it Symmetric and antisymmetric constitutive tensors forbi-isotropic and bi-anisotropic media}, arXiv : 2204.10460v2 [physics.class-ph].

\end{thebibliography}
\end{document}